\documentclass[11pt]{amsart}
\usepackage{amsmath}
\usepackage[margin=2.2cm]{geometry}
\usepackage{graphicx}
\usepackage{amssymb}
\usepackage{epstopdf}
\usepackage{color}
\usepackage{ulem}

\usepackage{amsaddr}

\newcommand{\CR}[1]{{\color{black}{#1}}} 
\newcommand{\CB}[1]{{\color{black}{#1}}} 

\newcommand{\CE}[1]{{{}}}

\newcommand{\TP}{\mathrm{TP}}
\newcommand{\TN}{\mathrm{TN}}
\newcommand{\FP}{\mathrm{FP}}
\newcommand{\FN}{\mathrm{FN}}

\title{ Sparse causality network retrieval from short time series } 
\author{Tomaso Aste$^{1,2,3}$ and T. Di Matteo$^{1,2,3,4}$}
\address{$^1$ Department of Computer Science, UCL, London, UK\\
$^{2}$ UCL Centre for Blockchain Technologies, UCL, London, UK\\
$^{3}$ Systemic Risk Centre, London School of Economics and Political Sciences, London, UK\\
$^{4}$ Department of Mathematics King's College London, London, UK }
\date{\today}                                           
\begin{document}
\begin{abstract}
We investigate how efficiently a known underlying \CR{sparse} causality structure of a simulated multivariate \CR{linear} process can be retrieved from the analysis of time-series \CR{of short lengths}. 
Causality is quantified from conditional transfer entropy and the network is constructed by retaining only the statistically validated contributions. 
We compare results from three methodologies:  two commonly used regularization methods,  Glasso and ridge, and a newly introduced technique, LoGo, based on the combination of information filtering network and graphical modelling.
For these three methodologies we explore the regions of time series lengths and model-parameters where a significant fraction of true causality links is retrieved. 
We conclude that,  when time-series are short, with their lengths shorter than the number of variables,  sparse models are better suited to uncover  true causality links with LoGo retrieving the true causality network more accurately than Glasso and ridge. 
\end{abstract}

\maketitle
\section{Introduction}
Establishing causal relations between variables from observation of their behaviour in time is central to scientific investigation and it is at the core of data-science where these causal relations are  the basis for the construction of useful models and tools capable of prediction.
The capability to predict (future) outcomes from the analytics of (past) input data is crucial in modeling and it should be the main property to take into consideration in model selection, when  the validity and meaningfulness of a model is assessed. 
From an high-level perspective, we can say that the whole scientific method is constructed around a circular procedure consisting in observation, modelling, prediction and testing. In such a procedure, the accuracy of prediction is used as a selection tool between models. 
In addition, the principle of parsimony favours the simplest model when two models have similar predictive power.

The scientific method is the rational process that, for the last 400 years, has mostly contributed to scientific discoveries, technological progresses and the advancement of human knowledge. 
Machine learning and data-science are nowadays  pursuing the ambition to mechanize this discovery process by feeding machines with data and using different methodologies to build systems able to make models and predictions by themselves.
However, the automatisation of this process requires to identify, without the help of human intuition, the relevant variables and the relations between these variables out of a large quantity of data. 
Predictive models are methodologies,  systems or equations which identify and make use of such relations between  sets of variables in a way that  the knowledge about a set of variables provides information about the values of the other set of variables.
This problem is intrinsically high-dimensional with many input and output data. 
Any model that aims to explain the underlying system will involve  a number of elements which must be of the order of magnitude of the number of relevant relations between the system's variables. 
In complex systems, such as financial markets or the brain, prediction is probabilistic in nature and modeling concerns inferring the probability of the values of a set of variables given the values of another set.
This requires the estimation of the joint probability of all variables in the system and, in complex systems,  the number of variables with potential macroscopic effects on the whole system is very large. 
This poses a great challenge for the model construction/selection and its parameter estimation because  the number of relations between variables scales with -at least- the square of the number of variables but, \CR{for a given fix} observation window, the amount of information gathered from such variables scales -at most- linearly with the number of variables \cite{bruckstein2009sparse,theodoridis2012sparsity}. 

For instance, a linear model for a system with $p$ variables  requires the estimation  from observation of $p(p+1)/2$  parameters (the distinct elements of the covariance matrix).
In order to estimate $\mathcal O(p^2)$  parameters one needs a comparable number of observations requiring  time series of length $q \sim p$ or larger to gather a sufficient information content from a number of observations which scales as $p \times q \sim \mathcal O(p^2)$.
However, the number of parameters in the model can be reduced by considering only $\mathcal O(p)$  out of the $\mathcal O(p^2)$ relations between the variables reducing in this way the required time series  length to $\mathcal O(p)$.
Such models with reduced numbers of parameters are referred in the literature as sparse models.
In this paper we consider two instances of linear sparse modelling: Glasso \cite{tibshirani1996} which penalizes non-zero parameters by introducing a $\ell_1$ norm penalization and LoGo \cite{LoGo16} which reduces the inference network to an $\mathcal O(p)$ number of links selected by using information filtering networks  \cite{asteetal2005,tumminelloetal2005,TMFG}. 
The results from these two sparse models are compared with the $\ell_2$ norm penalization (non-sparse) ridge  model \cite{tikhonov1963solution,hoerl1970ridge}.

{
This paper is an exploratory attempt to map the parameter-regions of time series length, number of variables, penalization parameters and kinds of models to define the boundaries where probabilistic models can be reasonably constructed from the analytics of observation data. 
In particular, we investigate empirically, \CR{by means of a linear autoregressive model with sparse inference structure,} the true \CR{causality} link retrieval performances in the region of short time-series and large number of variables which is the most critical region -- and the most interesting -- in many practical cases. 
\CR{Causality is defined in information theoretic sense as a significant  reduction on uncertainty over the present values of a given variable provided by the knowledge of the past values of another variable obtained in excess to the knowledge provided by the past of the variable itself and --in the conditional case--  the past of all other variables \cite{zaremba2014measures}. We measure such information by using transfer entropy and, within the present linear modelling, this coincides with the concept of Granger causality and conditional Granger causality \cite{schreiber2000measuring}. 
The use of transfer entropy has the advantage of being a concept directly extensible to non-linear modelling. 
However, non-linearity is not tackled within the present paper.
Linear models with multivariate normal distributions have the unique advantage that causality and partial correlations are directly linked, largely simplifying the computation of transfer entropy and directly mapping the problem into the sparse inverse covariance problem \cite{tibshirani1996,LoGo16}.
}

Results are reported for artificially generated time series from an autoregressive model of $p=100$ variables and time series lengths $q$ between 10 and 20,000   data points. 
Robustness of the results has been verified over a wider range of $p$ from 20 to 200 variables.
Our results demonstrate that sparse models are superior in retrieving the  true causality structure for short time series. 
Interestingly, this is despite considerable inaccuracies in the inference network of these sparse models.
We indeed observe that statistical validation of causality is crucial in identifying the true causal links, and this identification is highly enhanced in sparse models.
}

The paper is structured as follows. 
In section \ref{s.definitions} we briefly review  the basic concepts of mutual information and conditional transfer entropy and their estimation from data that will then be used in the rest of the paper. 
We also introduce the concepts of sparse inverse covariance, inference network and causality networks.
Section \ref{s.causalNet} concerns the retrieval of causality network from the computation and statistical validation of conditional transfer entropy.
Results are reported in Section \ref{s.results} where the retrieval of the true causality network from the analytics of time series from an autoregressive process of $p=100$ variables is discussed.
Conclusions and perspectives are given in Section \ref{s.conclusions}.

\section{Estimation of conditional transfer entropy from data} 
\label{s.conditionslSigma}\label{s.definitions} 

In this paper causality is quantified  by means of  statistically-validated transfer entropy. 
Transfer entropy  $T(\mathbf{ Z_i} \rightarrow \mathbf{ Z_j})$  quantifies the amount of uncertainty on a random variable, $\mathbf{ Z_j}$, explained by {the past} of another variable, $\mathbf{Z_i}$ conditioned to the knowledge about the past of  $\mathbf{ Z_j}$ itself.
Conditional transfer entropy, $T(\mathbf{ Z_i} \rightarrow \mathbf{ Z_j} | \mathbf W)$, includes an extra condition also to a set variables $\mathbf W$. 
These quantities are introduced in details in Appendix \ref{cTE} (see also \cite{shannon2001mathematical,schreiber2000measuring,anderson1984multivariate}). 
Let us here just report the main expression for the conditional transfer entropy that we shall use in this paper:
\begin{align}\label{General_cTE}
T(\mathbf{ Z_i} \rightarrow \mathbf{ Z_j}|\mathbf{ W}) & {}= 
H(\mathbf Z_{\mathbf j,t} | \{\mathbf Z_{\mathbf j,t}^{lag},\mathbf{ W}_{t}\})
- 
H(\mathbf Z_{\mathbf j,t} | \{\mathbf Z_{\mathbf i,t}^{lag},\mathbf Z_{\mathbf j,t}^{lag},\mathbf{ W}_{t}\})	\;\;.	
\end{align}
Where $H(.|.)$ is the conditional entropy,  $\mathbf Z_{\mathbf j,t}$ is a random variable at time $t$, whereas $\mathbf Z_{\mathbf i,t}^{lag} =\{ \mathbf Z_{\mathbf i,t-1},...,\mathbf Z_{\mathbf i,t-\tau}\}$ is the lagged set of random variable `$\mathbf i$' considering previous times $t-1...t-\tau$ and $\mathbf{ W}_{t}$ are all other variables and their lags (see Appendix \ref{cTE}, Eq.\ref{General_cTE1}).

In this paper we use Shannon entropy and restrict to linear modeling with multivariate normal setting (see Appendix \ref{Shannon}).
In this context the conditional transfer entropy can be expressed in terms of the determinants of conditional covariances  $\mbox{det}( \mathbf \Sigma(.|.))$ (see Eq.\ref{H2} in  Appendix \ref{Shannon}):
\begin{align}\label{General_cTE_Sigma}
T(\mathbf{ Z_i} \rightarrow \mathbf{ Z_j}|\mathbf{ W}) & {}= 
\frac12 \log \mbox{det}\!\left( \mathbf \Sigma(\mathbf Z_{\mathbf j,t} | \{\mathbf Z_{\mathbf j,t}^{lag},\mathbf{ W}_{t}\})\right)
- 
\frac12 \log \mbox{det}\!\left( \mathbf \Sigma(\mathbf Z_{\mathbf j,t} | \{\mathbf Z_{\mathbf j,t}^{lag},\mathbf Z_{\mathbf i,t}^{lag},\mathbf{ W}_{t}\})\right)	\;\;.	
\end{align}
 
Conditional covariances can be conveniently computed in terms of the inverse covariance of the whole set of variables 
$\mathbf Z_t = \{ \mathbf  Z_{k,t},\mathbf  Z_{k,t-1},...\mathbf  Z_{k,t-\tau}\}_{k=1}^p  \in \mathbb R^{p\times(\tau+1)}$ 
(see  Appendix \ref{InverseCovJandSigma}). 
Such inverse covariance matrix, $\mathbf J$,  represents the structure of conditional dependencies among all couples of variables  in the system and their lags.
Each sub-part of $\mathbf J$ is associated with the conditional covariances of the variables in that part with respect to all others. 
In terms of $\mathbf J$, the expression for the conditional transfer entropy becomes:
\begin{align}\label{TE20}
T(\mathbf{ Z_i} \rightarrow \mathbf{ Z_j}|\mathbf{ W})  =  - \frac12 \log\mbox{det}\! \left(\mathbf {J_{1,1}} - \mathbf {J_{1,2}} (\mathbf {J_{2,2}})^{-1} \mathbf {J_{2,1}})\right)+ \frac12 \log \mbox{det}\! \left (\mathbf{ J_{1,1} }\right ) \;\;.
 \end{align}
where the indices `$\mathbf 1$' and `$\mathbf 2$' refer to sub-matrices of $\mathbf J$ respectively associated with the variables $ \mathbf Z_{\mathbf j,t}$ and $\mathbf Z_{\mathbf i,t}^{lag}$.

\CB{
\subsection{Causality and inference networks}
The inverse covariance $\mathbf J$, also known as precision matrix, represents the structure of conditional dependencies. 
If we interpret the structure of $\mathbf J$  as a network, where nodes are the variables and non-zero entries correspond to edges of the network, then we shall see that any two sub-sets of nodes that are not directly connected by one or more edges are conditionally-independent. 
Condition is with respect to all other variables.

Links between variables at different lags are associated with causality with direction going from larger  to smaller lags. 
The network becomes therefore a directed graph.
In such a network  entropies can be associated with nodes, conditional mutual information can be associated with edges between variables with the same lag and conditional transfer entropy can be associated to edges between variables with  different lags.
A nice property of this mapping of information measures with directed networks is that there is a simple way to aggregate information which is directly associated with topological properties of the network.
Entropy, mutual information and transfer entropies can be defined for any aggregated sub set of nodes with their values  directly associated to the presence,  direction and weight of network edges between these sub-parts.

Non-zero transfer entropies indicating --for instance-- variable $\mathbf i$ causing variable $\mathbf j$ are associated with some non-zero entries in the inverse covariance matrix $\mathbf J$ between lagged variables $\mathbf i$ (i.e. $\mathbf{ Z_{i}}_{,t-\tau}$, with $\tau >0$) and variable $\mathbf j$ (i.e. $\mathbf{ Z_{j}}_{,t}$)).
\CR{In linear models, }these non-zero entries define the \CR{estimated} {\it inference network}.
However, not all edges in this inference network correspond to transfer entropies that are significantly different from zero. 
To extract the structure of the {\it causality network} we shall retain only the edges in the  inference network which correspond to statistically validated transfer entropies.
}

Conditioning eliminates the effect of the other variables retaining only the exclusive contribution from the two variables in consideration. 
This should provide estimations of transfer entropy that are less affected by spurious effects from other variables.
On the other hand, conditioning in itself can introduce spurious effects, indeed two independent variables can become dependent due to conditioning  \cite{anderson1984multivariate}.
In this paper we explore two extreme conditioning cases: i) condition to all other variables and their lags; ii) unconditioned. 

{
In principle, one would like to identify the maximal value of $T(\mathbf{ Z_i} \rightarrow \mathbf{ Z_j} |  \mathbf W)$ over all lags and all possible conditionings $\mathbf W$. 
However, the use of multiple lags and conditionings increases the dimensionality of the problem making estimation of transfer entropy very hard especially when only a limited amount of measurements is available (i.e. short time-series). 
This is because the calculation of the conditional covariance requires the estimation of the inverse covariance of the whole set of variables and such an estimation is strongly affected by noise and uncertainty. 
Therefore, a standard approach is to reduce the number of variables and lags to keep dimensionality low and estimate  conditional covariances with appropriate penalizers \cite{tikhonov1963solution,hoerl1970ridge,tibshirani1996,friedmanetal2008}. 
An alternative approach is to invert the covariance matrix only locally on  low dimensional sub-sets of variables selected by  using information filtering networks \cite{asteetal2005,tumminelloetal2005,TMFG} and then reconstruct the global inversion  by means of the LoGo approach \cite{LoGo16}.
Let us here briefly account for these two approaches. }

\subsection{Penalized inversions}
The estimate of the  inverse covariance is a challenging task to which a large body of literature has been dedicated \cite{friedman2008sparse}.
From an intuitive perspective, one can say that the problem lies in the fact that uncertainty is associated with nearly zero eigenvalues of the covariance matrix. 
Variations in these small eigenvalues have relatively small effects on the entries of the covariance matrix itself but have major effects on the estimation of its inverse.  
Indeed small fluctuations of small values can yield to unbounded contributions to the inverse.  
A way to cure  such near-singular matrices is by adding finite positive terms to the diagonal which move the eigenvalues away from zero:
$\hat {\mathbf J} = \left((1-\gamma) \mathbf S +\gamma \mathbf I_N \right)^{-1}$, where $\mathbf S = \mbox{Cov}(\mathbf Z)$ is the covariance matrix of the set of variables $\mathbf Z \in \mathbb R^N$ estimated from data and  $\mathbf I_{N}\in \mathbb R^{N\times N}$ is the identity matrix (where $N=p\times(\tau+1)$, see later).
This is what is performed in the so-called ridge regression \cite{hoerl1970ridge}, also known as shrinkage mean-square-error estimator \cite{gruber1998improving} or Tikhonov regularization \cite{tikhonov1963solution}.
The effect of the additional positive diagonal elements is quivalent to compute the inverse covariance which maximizes the log-likelihood: $\log \mbox{det}( \hat {\mathbf  J}) - \mbox{tr}(\mathbf S\hat{\mathbf J}) - \gamma  ||\hat {\mathbf J} ||_2$, where the last term penalizes large off-diagonal coefficients in the inverse covariance with a $\ell_2$ norm penalization \cite{witten2009covariance}.  
The regularizer parameter $\gamma$ tunes the strength of this penalization.
This regularization is very simple and effective. 
However, with this method insignificant elements in the  precision  matrix are penalized toward small values  but they are never set to zero.
By using  instead  $\ell_1$ norm penalization: $\log \mbox{det}( \hat {\mathbf  J}) -  \mbox{tr}(\mathbf S\hat{\mathbf J}) - \gamma  ||\hat{\mathbf J} ||_1$, insignificant   elements are forced to zero leading to a sparse  inverse covariance. 
This is the so-called lasso regularization \cite{tibshirani1996,meinshausenbuehlmann2006,friedmanetal2008}.
The advantage of a sparse  inverse covariance consists in the provision of a network \CR{representing a conditional dependency  structure.} Indeed, let us recall that \CR{in linear models} zero entries in the  inverse covariance are associated with couples of non-conditionally dependent variables.

\subsection{Information filtering network approach:  LoGo}
An alternative approach to obtain sparse  inverse covariance is by using  information filtering networks generated by keeping the elements that  contribute most to the covariance by means of a greedy process. 
%How to construct such networks is described in \cite{TMFG,LoGo16}.
\CR{This approach, named LoGo, proceeds by first constructing a chordal information-filtering graph such as a Maximum Spanning Tree (MST) \cite{kruskal1956,mantegna1999} or a Triangulated Maximally Filtered Graph (TMFG) \cite{TMFG}. 
These graphs are build by retaining edges that maximally contribute to a given gain function which, in this case, is the log-likelihood or --more simply-- the sum of the squared correlation coefficients \cite{asteetal2005,tumminelloetal2005,TMFG}.
Then, this chordal structure is interpreted as the inference structure of the joint probability distribution function with non-zero conditional dependency only between variables that are directly connected by an edge. 
On this structure the sparse inverse covariance is computed in such a way to preserve the values of the correlation coefficients between couples of variables that are directly connected with an information-filtering graph edge.    
The main advantage} of this approach is that inversion is performed at local level on a small subsets of variables and then the global inverse is reconstructed by joining the local parts  through the information filtering network.
Because of this Local-Global construction this method is named LoGo.
It has been shown that LoGo method yields to statistically significant sparse  precision matrices that outperform the ones with the same sparsity computed with lasso method \cite{LoGo16}.

\section{Causality network reterival} \label{s.causalNet}

\subsection{Simulated multivariate autoregressive linear process}
In order to be able to test if  causality measures can retrieve  the true causality network in the underlying process, we generated artificial multivariate normal time series with known sparse causality structure by using the following autoregressive multivariate linear process \cite{hamilton1994time}:
\begin{align}\label{Process}
\mathbf  Z_t = \sum_{\lambda=1}^\tau \mathbf A_\lambda \mathbf  Z_{t-\lambda}  + \mathbf U_t 
\end{align}
where   $\mathbf A_\lambda \in \mathbb R^{p \times p}$ are  matrices with random entries drawn from a normal distribution. 
The matrices are made upper diagonal (diagonal included) by putting to zero all lower diagonal coefficients and made sparse by keeping only a  $\mathcal O(p)$ total number of entries different from zero in the upper and diagonal part. 
$\mathbf U_t \in \mathbb R^p$ are random normally distributed uncorrelated variables.  
This process produces autocorrelated, cross-correlated and causally dependent time series. 
We chose it because it is among the simplest processes that can generate this kind of structured datasets.
The dependency and causality structure is determined by the non-zero entries of the matrices $\mathbf A_\lambda$.
The upper-triangular structure of these matrices simplify the causality structure eliminating causality cycles. 
Their sparsity reduces dependency and causality interactions among variables.  
The process is made autoregressive and stationary by keeping the  eigenvalues of $\mathbf A_\lambda$ all smaller than one in absolute value.
For the tests we used $\tau=5$, $p=100$ and sparsity is enforced to have a number of links approximately equal to $p$.
We reconstructed the network from time series of different lengths $q$ between 5 to 20,000 points.  
To test statistical reliability the process was repeated 100 times with every time a different set of  randomly generated matrices  $\mathbf A_\lambda$.
We verify that the results are robust and consistent by varying sample sizes from $p=20$ to $200$, by changing sparsity with number of links from $0.5 p$ to $5 p$ and for $\tau$ from 1 to 10. 
\CR{We verified that the presence of isolated nodes or highly connected hub nodes does not affect results significantly.}

\subsection{Causality and inference network retrieval}
We tested the agreement between the causality structure of the underlying process and the one inferred from the analysis of  $p$ time-series of different lengths $q$, $\mathbf  Z_t \in \mathbb R^p$ with $t=1..q$, generated by using Eq.\ref{Process}.
We have $p$ different variables and $\tau$ lags.
The dimensionality of the problem is therefore $N = p \times (\tau+1)$ variables at all lags including zero. 

To estimate the  inference and causality networks we started by computing the  inverse covariance, $\mathbf J   \in  \mathbb R^{ N \times N}$, for all variables at all lags  $\mathbf  Z \in  \mathbb R^{ N \times q}$ by using the following three different estimation methods:
\begin{itemize}
\item[1)] $\ell_1$  norm penalization (Glasso \cite{friedmanetal2008}); 
\item[2)] $\ell_2$ norm penalization (ridge \cite{tikhonov1963solution}); 
\item[3)] information filtering network  (LoGo \cite{LoGo16}).
\end{itemize}

We  retrieved the inference network by looking at all couples of variables, with indices ${\mathbf i} \in[1,..,p]$ and ${\mathbf j} \in[1,..,p]$, which have non-zero entries in the inverse covariance matrix $\mathbf J$ between the lagged set of $\mathbf j$ and the non-lagged $\mathbf i$. 
Clearly, for the ridge method the result is a complete graph but for the Glasso  and LoGo the results are sparse networks with edges corresponding to non-zero conditional transfer entropies between variables $\mathbf i$ and $\mathbf j$.
For the LoGo calculation we make use of the regularizer parameter as a local shrinkage factor to improve the local inversion of the covariance of the 4-cliques and triangular separators (see \cite{LoGo16}).

We then estimated   transfer entropy between couples of variables, ${\mathbf i}  \rightarrow  {\mathbf j}$ conditioned to  all other variables in the system. 
This is obtained by estimating of the inverse covariance matrix (indicated with an `hat' symbol) by using Eq.\ref{TE20b} (see Appendix \ref{s.CTE}) with:
\begin{align}
\mathbf{Z_1} &=  \mathbf{  Z_j}_{,t}\\ \nonumber
\mathbf{Z_2} &=\{\mathbf{  Z_i}_{,t-1}...\mathbf{  Z_i}_{,t-\tau}\}\\ \nonumber
\mathbf{Z_3} &= \{\mathbf{  Z_j}_{,t-1}...\mathbf{  Z_j}_{,t-\tau},\mathbf{ W}\} \;.
 \end{align}
With $\mathbf{ W}$ a conditioning to all variables $\mathbf{Z}$ except $\mathbf{Z_1},\mathbf{Z_2}$ and $\{\mathbf{  Z_j}_{,t-1}...\mathbf{  Z_j}_{,t-\tau}\}$.
The result is a $p \times p$ matrix of conditional transfer entropies $T(\mathbf{  Z_i}_{,t} \rightarrow  \mathbf{  Z_j}_{,t})$.
Finally, to retrieve the causality network we retained the network of  statistically validated   conditional transfer entropies only.
Statistical validation was performed as follows.

\subsection{Statistical validation of  causality}
{Statistical validation has been performed from likelihood ratio statistical test. Indeed,  entropy and likelihood are intimately related: entropy measures uncertainty and likelihood measures the reduction in uncertainty provided by the model.
Specifically, the Shannon entropy associated with a set of random variables, $\mathbf{ Z_i}$, with probability distribution $p(\mathbf{ Z_i})$   is $H(\mathbf{ Z_i})  =  - \mathbb E [ \log p(\mathbf{ Z_i}) ]$ (Eq.\ref{entropy}) whereas the log-likelihood for the model $\hat p(\mathbf{ Z_i})$ associated with a set of independent observations $\mathbf{ \hat Z}_{i,t}$ with $t=1..q$ is  $\log \mathcal L(\mathbf{\hat Z}_{\mathbf i})  = \sum_{t=1}^q \log \hat p(\mathbf{\hat Z}_{i,t})$ which can be written as $\log \mathcal L(\mathbf{\hat Z}_{\mathbf i})  = q \mathbb E_{\hat p} [ \log \hat p(\mathbf{ Z_i}) ]$. 
Note that $q$ is the total available number of observations which, in practice, is the length of the time-series minus the maximum number of lags.
It is evident from these expressions that  entropy and the log-likelihood are strictly related trough this link might be non-trivial.
In the case of linear modeling this connection is quite evident because  the entropy estimate is $H =  \frac 12 (- \log |\mathbf{\hat J} | + p \log (2\pi) + p)$ and the log-likelihood is $\log \mathcal L = \frac q2 ( \log |\mathbf{\hat J} | - Tr( \mathbf{\hat \Sigma} \mathbf{\hat J} ) - p \log (2\pi))$. 
For the three models we study in this paper we have $Tr( \mathbf{\hat \Sigma} \mathbf{\hat J} ) =p$ and therefore  the log-likelihood is equal to $q$ times the opposite of the entropy estimate.
Transfer entropy, or conditional transfer entropy, are differences between two entropies: the one of a set of variables conditioned to their own past minus the one conditioned also to {the past} of another variable. 
This, in turns, is the difference of the unitary log-likelihood of two models and therefore it is the logarithm of a likelihood ratio.
As Wilks pointed out \cite{wilks1938large,vuong1989likelihood} the null distribution of such model is asymptotically quite universal.
Following the likelihood ratio formalism, we have $\lambda=q T$ and the probability of observing a transfer entropy larger than $T$, estimated under null hypothesis, is given by $p_v  \sim 1-\chi_c^2(r qT,d)$ with $r\simeq 2$ and $\chi_c^2$ the chi-square the cumulative distribution function with $d$ degrees of freedom which are the difference between the number of parameters in the two models. 
In our case the two models have respectively $\tau  (p_j^2+1)$ and $\tau(p_j^2+1) + \tau  (p_j \, p_i)$ parameters. 
}

\subsection{Statistical validation of  the network}
The procedures described in the previous two subsections produce the inference network and causality network. 
Such networks are then compared with the known underlying network of true causalities in the underlying process which is defined by the non-zero elements in the matrices $A_\lambda$  (see Eq.\ref{Process}).
The overlapping between the retrieved  links in the inference  or causality networks with the ones   in the true network underlying the process is an indication of a discovery of a true causality relation.
However some  discoveries can be obtained just by chance or some methodologies might discover more links only because they produce denser networks. 
We therefore tested the hypothesis that the matching links in the retrieved networks are not obtained just by chances by computing the null-hypothesis probability to obtain the same or a larger number of matches randomly.
Such probability is given by the conjugate cumulative hypergeometric distribution for a number equal  or larger than $\TP$ of `true positive' matching causality links between an inferred network of $n$ links and a process network of $K$ true causality links, from a population of $p^2-p$ possible links:
\begin{align}\label{HyperTest}
P(X \ge \TP | n , K , p)= 1 -  \sum_{k=0}^{\TP-1} {\frac {{\binom {K}{k}}{\binom {p^2-p -K}{n-k}}}{\binom {p^2-p}{n}}} \;\;\;.
\end{align}
Small values of $P$ indicate that the retrieved $\TP$ links out of $K$ are unlikely to be found by randomly picking $n$ edges from $p^2-p$ possibilities.
Note that in the  confusion matrix notation \cite{swets2014signal} we have \CR{ $n=\TP+\FP$ and $K = \TP+\FN$ }with $\TP$ number of true positives,  $\FP$ number of false positives,  $\FN$ number of false negatives and $\TN$ number of true negatives.  
\CR{The total number of `negative' (unlinked couples of vertices) in the true model is instead $m=\FP+\TN$.}

\section{Results}\label{s.results}

\subsection{Computation and validation of conditional transfer entropies}
By using Eq.\ref{Process} we generated 100 multivariate autoregressive processes with known causality structures.
We here report results for $p=100$  but analogous outcomes were observed for dimensionalities between $p=20$ and $200$ variables.
Conditional transfer entropies between all couples of variables, conditioned to all other variables in the system, were computed  by estimating the inverse covariances by using tree methodologies, ridge, lasso and LoGo and applying Eq.\ref{TE20}.
Conditional transfer entropies were  statistically  validated with respect to null hypothesis (no causality) at $p_v=1\%$ p-value. Results for Bonferroni adjusted p-value at 1\% (i.e. $p_v = 0.01/(p^2-p) \sim 10^{-6}$ for $p=100$) are reported in Appendix \ref{BonferroniValidated}.
We also tested other values of $p_v$ from $10^{-8}$ to 0.1 obtaining consistent results.
We observe that small $p_v$  reduce the number of validated causality links but increase the chance that these links match with the true network in the process.  
Conversely large values of $p_v$ increase the numbers of mismatched links but also of the true links discoveries. 
Let us note that here we use $p_v$  as a thresholding criteria and we are not claiming any evidence of statistical significance of the causality. 
We assess the goodness of this choice a-posteriori by comparing the resulting causality network with the known causality network of the process.

\begin{center}
	\begin{figure}[t]
	\includegraphics[width=0.80\textwidth]{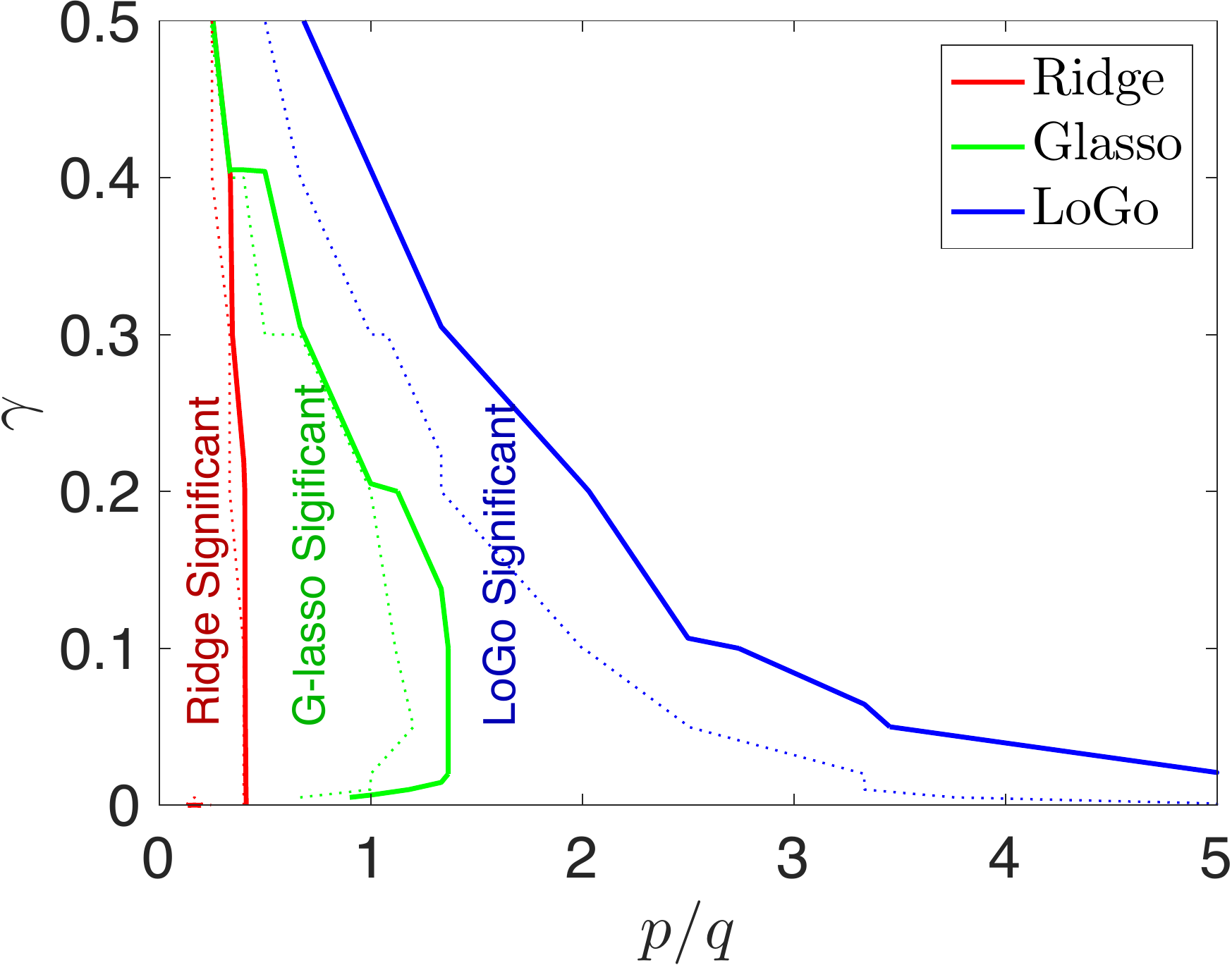}
		\caption{
		{\bf Regions in the $p/q$-$\gamma$ space where causality networks for the three models are statistically significant.} 
		The significance regions are all at the left of the corresponding lines.
		Tick line reports the boundary $P< 0.05$ (Eq.\ref{HyperTest}) and dotted lines indicate $P< 10^{-8}$ significance levels ($P$ is averaged over  100 processes).
		The plots refer to $p=100$ and report the region where the causality network are all significant for 100 processes. 		}
		\label{f.significance}
	\end{figure}
\end{center}

\begin{center}
	\begin{figure}[t]
	\includegraphics[width=0.80\textwidth]{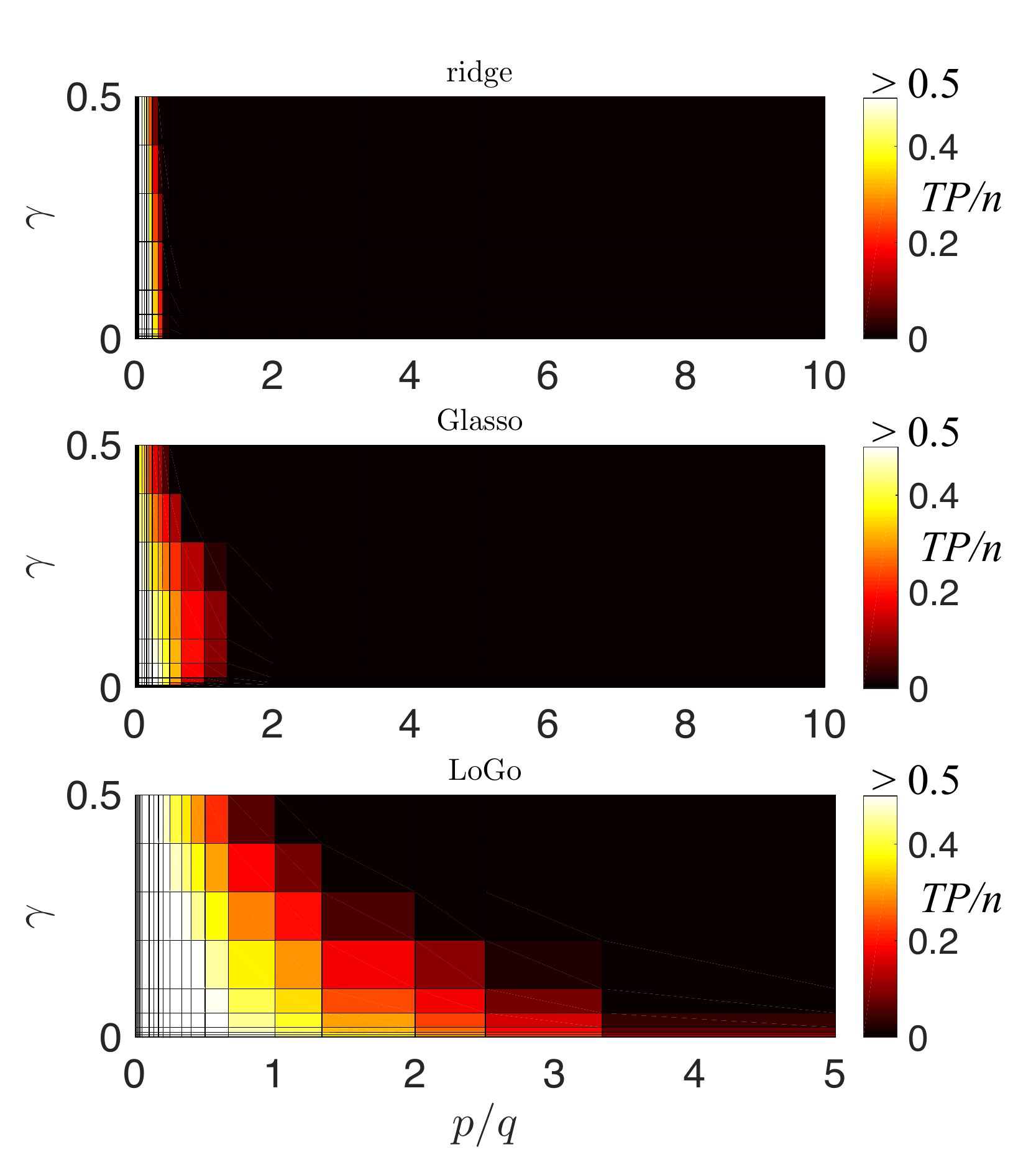}
		\caption{
		{\bf True positive rate: fraction of retrieved true causality links ($\TP$) with respect to the total number of links in the process (n).}
		The three panels refer to ridges,  Glasso and  LoGo (top, central and bottom).
		Data are average fractions over 100 processes.
		}
		\label{f.RegionFractionRetreived}
	\end{figure}
\end{center}

\subsection{Statistical significance of the recovered causality network.}
Results for the contour frontiers of significant causality links for the three models are reported in Fig.\ref{f.significance} for a range of time series with lengths $q$ between 10 and 20,000 and regularizer parameters $\gamma$ between $10^{-8}$ and $0.5$.
Statistical significance is computed by using Eq.\ref{HyperTest} and results are reported for both  $P < 0.05$ and $P < 10^{-8}$ (continuous and dotted lines respectively).
As one can see, the overall behaviours for the three methodologies are little affected by the  threshold on $P$.
We observe that LoGo significance region extends well beyond the Glasso and ridge regions.

The value of the regularizer parameter $\gamma$ affects the results for the three  models in a different way. 
Glasso has a region in the plane $\gamma-p/q$ where it has best performances (in this case it appears to be around $\gamma \simeq 0.1$ and $p/q \simeq 2.5$). 
Ridge  appears instead to be little affected with mostly constant performances across the range of $\gamma$.
LoGo has best performances for small, even infinitesimal, values of  $\gamma$. 
Indeed, differently from Glasso in this case  $\gamma$ does not control sparsity but instead acts as local shrinkage parameter.
Very small values can be useful in some particular cases to reduce the effect of noise but large values have only the effect to reduce information.

\begin{table}[!h]
\caption{
{\bf Causality network validation.} Comparison between fraction of true positive  (${TP/n}$) and fraction of false positive  ($FP/n$), statistically validated, causality links for the three models and different time-series lengths.
The table reports only the case for the parameter $\gamma=0.1$. 
Statistical validation of conditional transfer entropy is at $p_v=1\%$ p-value. 
Note that LoGo can perform better than reported in this table for smaller values of $\gamma$ (see Figs.\ref{f.significance} and \ref{f.RegionFractionRetreived}). 
}
\begin{center}
\begin{tabular}{@{\extracolsep{4pt}}llllllllll@{}}
\hline
q & 10 & 20 & 30 & 50 & 200 & 300 & 1000 & 20000  \\ \hline 
ridge TP/n  & 0.00  & 0.00  & 0.00  & 0.00  & 0.23$^{**}$  & 0.49$^{**}$  & 0.76$^{**}$  & 0.93$^{**}$ \\ 
ridge FP/n  & 0.00  & 0.00  & 0.00  & 0.00  & 0.00  & 0.10  & 0.65  & 1.06 \\ \hline 
Glasso TP/n  & 0.00  & 0.00  & 0.00  & 0.13$^{**}$  & 0.48$^{**}$  & 0.53$^{**}$  & 0.62$^{**}$  & 0.74$^{**}$ \\ 
Glasso FP/n  & 0.00  & 0.00  & 0.00  & 0.00  & 0.06  & 0.10  & 0.23  & 0.54 \\ \hline 
LoGo TP/n  & 0.00  & 0.08$^*$  & 0.21$^{**}$  & 0.37$^{**}$  & 0.61$^{**}$  & 0.65$^{**}$  & 0.75$^{**}$  & 0.90$^{**}$ \\ 
LoGo FP/n  & 0.00  & 0.00  & 0.00  & 0.01  & 0.06  & 0.08  & 0.15  & 0.34 \\ \hline 
&& $^{*}$  $P < 0.05$;& $^{**}$   $P < 10^{-8}$ &&&\\
\end{tabular}\end{center}
\label{tab:fraction}
\end{table}

\subsection{Causality links retrieval}
Once identified the parameter-regions where the retrieved causality links are statistically significant, we also measured  the fraction of true links retrieved.
\CR{Indeed, given that the true underlying causality network is sparse, one could do significantly better than random by discovering only a few true positives.
Instead, from any practical perspective we aim to discover a significant fraction of the edges. 
Figure \ref{f.RegionFractionRetreived} shows that  the fraction of causality links correctly discovered (true positive, $\TP$) with respect to the total number of  causality links in the process ($n$) is indeed large reaching values above 50\%.} 
This is the so-called true positive rate or sensitivity, which takes values between 0 (no links discovered) and 1 (all links discovered).
Reported values are averages over 100 processes. 
We observe that the region  with discovering of 10\% or more true causality links greatly overlaps with the statistical validity region of Fig.\ref{f.significance}. 

We note that when the observation time becomes long, $p/q  \lessapprox 0.25$, ridge discovery rate  becomes larger than LoGo. 
However, statistical significance is still inferior to  LoGo, indeed the ridge network becomes dense when $q$ increases and the larger discovery rate of true causality links is also accompanied by a larger rate of false links incorrectly identified (false positive $\FP$).

The fraction of false positives with respect to the total number of  causality links in the process ($n$) are reported in Table~\ref{tab:fraction} together with the true positive rate for comparison.
This number can reach values larger than one because the process is sparse and there are much more possibilities to randomly chose false links than true links. 
\CR{Note this is not the false positive rate, which instead is $\FP/m$, and cannot be larger than one.}
Consistently with Fig.\ref{f.significance} we observe that, for short time series, up to $p/q \sim 0.5$  the sparse models have better capability to identify true causality links and to discard the false ones with LoGo being superior to Glasso.
Remarkably, LoGo can identify a  significant fraction of causality links already from time-series with lengths of 30 data-points only.
P-value significances, reported in the table with one or two stars indicate when all values of $P(X \ge \TP | n , K , p)$ from Eq.\ref{HyperTest} for all 100 processes have respectively  $P<0.05$ or $P<10^{-8}$.
Again we observe that  LoGo discovery rate region extends well beyond the Glasso and ridge regions.

\begin{table}[!h]
\caption{
{\bf Inference network validation: comparison between fraction of true positive (TP/n) and fraction of false positive  (FP/n).}
Data for ridge are only for comparison because it is a complete graph with all links present.
The table reports only the case for the parameter $\gamma=0.1$. 
}
\begin{center}
\begin{tabular}{@{\extracolsep{4pt}}llllllllll@{}}
\hline
q & 10 & 20 & 30 & 50 & 200 & 300 & 1000 & 20000  \\ \hline 
ridge TP/n  & 1.00  & 1.00  & 1.00  & 1.00  & 1.00  & 1.00  & 1.00  & 1.00 \\ 
ridge FP/n  & 97.84  & 97.84  & 97.84  & 97.84  & 97.84  & 97.84  & 97.84  & 97.84 \\ \hline 
Glasso TP/n  & 0.61$^*$  & 0.74$^*$  & 0.79$^*$  & 0.85$^*$  & 0.87$^{**}$  & 0.84$^{**}$  & 0.80$^{**}$  & 0.80$^{**}$ \\ 
Glasso FP/n  & 28.39  & 38.11  & 45.79  & 53.58  & 40.61  & 26.60  & 1.54  & 0.92 \\ \hline 
LoGo TP/n  & 0.31$^*$  & 0.50$^{**}$  & 0.58$^{**}$  & 0.63$^{**}$  & 0.75$^{**}$  & 0.78$^{**}$  & 0.85$^{**}$  & 0.93$^{**}$ \\ 
LoGo FP/n  & 4.53  & 4.27  & 4.18  & 4.03  & 3.72  & 3.63  & 3.44  & 3.21 \\ \hline 
&& $^{*}$  $P < 0.05$; & $^{**}$  $P < 10^{-8}$ &&&\\
\end{tabular}\end{center}
\label{tab:fraction_Inference}
\end{table}

\begin{center}
	\begin{figure}[t]
	\includegraphics[width=0.45\textwidth]{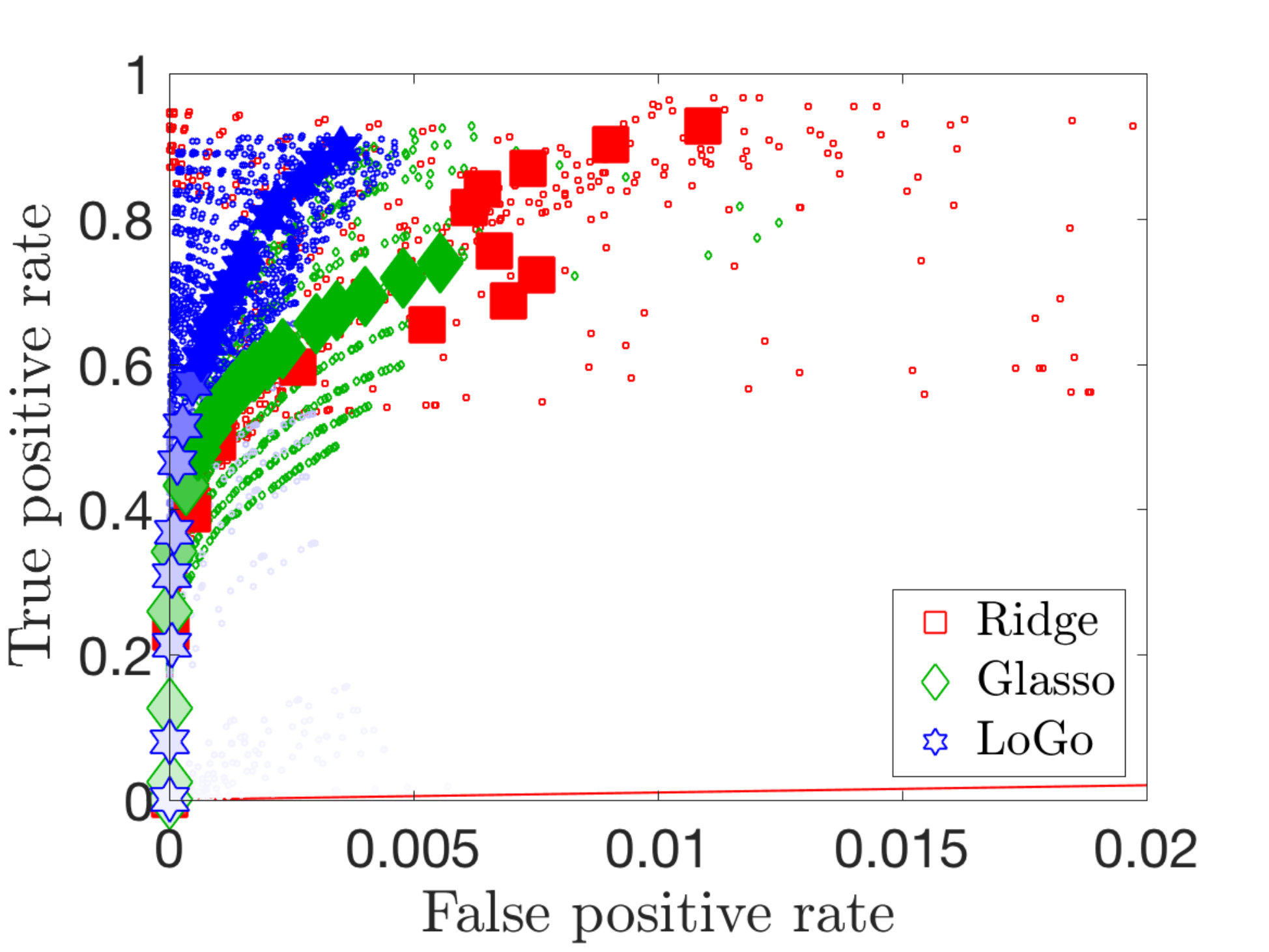}		
	\includegraphics[width=0.45\textwidth]{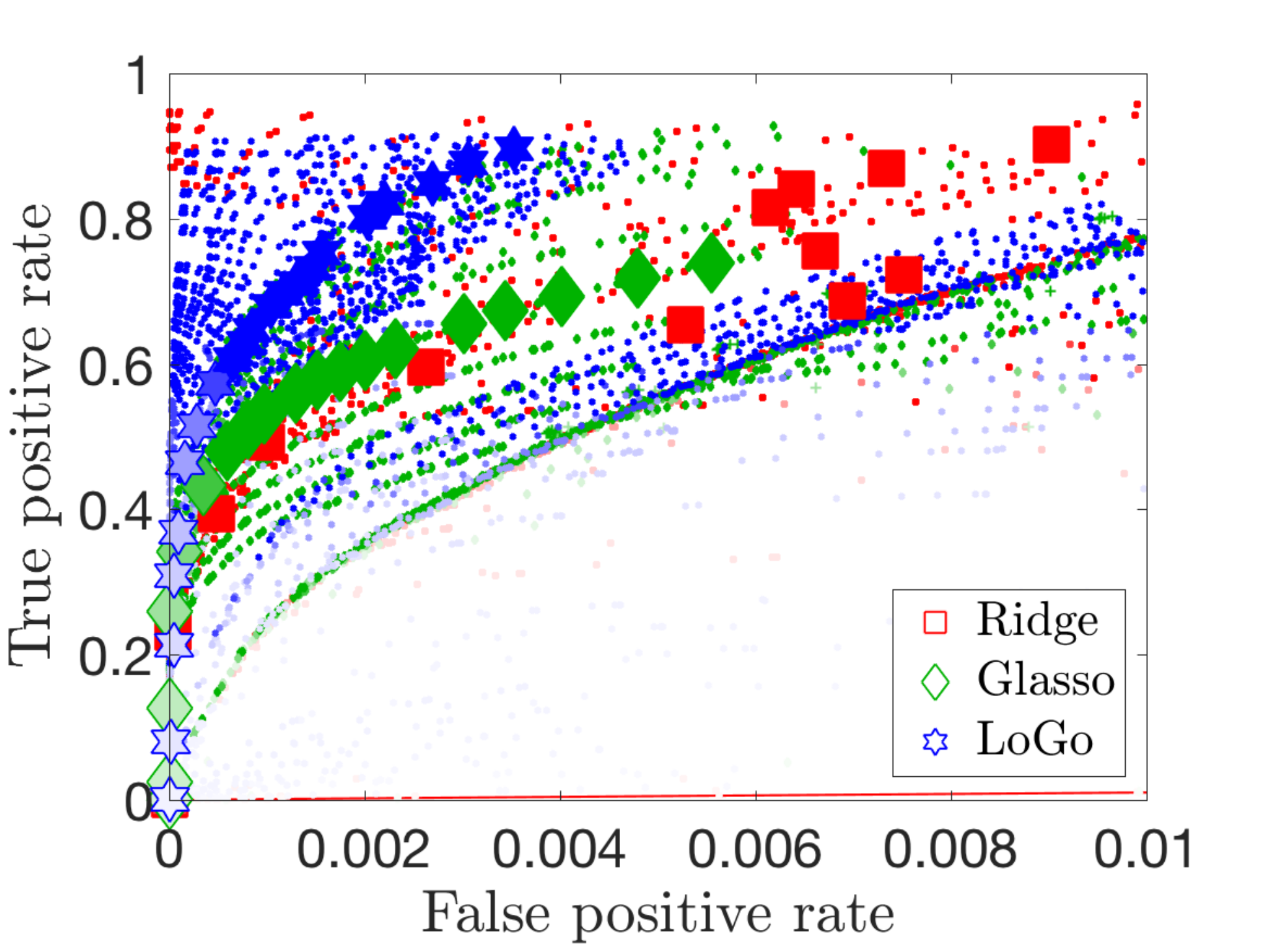}
	\caption{
		{\bf ROC values, for each model and each parameter combination.} X-axis  false positive rates  ($\FP/m$), y-axis true positive rates ($\TP/n$).
		Left and right figures are the same with X-axis expanded on the low values only for the right figure to better visualise the differences between the various models.
		Large symbols refer to $\gamma = 0.1$ and  validation at p-value $p_v=0.01$.
		Color intensity is proportional to time series length.
		Inference network results are  all outside the range of the plot. 
		Reported values are averages over 100 processes.	}
		\label{f.ROC}
	\end{figure}
\end{center}

\subsection{Inference network}
We have so far empirically demonstrated  that a significant part of the true causality network can be retrieved from the statistically validated network of conditional transfer entropies. 
Results depend on the choice of the threshold value of the $p_v$ at which null hypothesis is rejected.
We observed that lower $p_v$ are associated with network with fewer true positives but also fewer false positives and conversely larger $p_v$ yield to causality networks with larger true positives but also larger false positives.
Let us here report on the extreme case of the inference network which contains all causality channels with no validation.
For the ridge model this network is the complete graph with all variables connected to each-other.
Instead, for Glasso and LoGo the inference network is  sparse. 

Results are summarized in Table.\ref{tab:fraction_Inference}.
In terms of true positive rate we first notice that they are all larger than the ones in Table.\ref{tab:fraction}. 
Indeed, the network of statistically validated conditional transfer entropies is a sub-network of the  inference network.
On the other hand we notice that the false positive fraction is much larger than the ones in  Table.\ref{tab:fraction_Inference}.
Ridge network has a fraction of 1 because, in this case, the inference network is the complete graph.

Galsso also contains a very large number of false positives reaching even 55 times the number of  links in the true network and getting to  lower fractions only from long time-series with $q >1000$.
These numbers also indicate that Galsso networks are not sparse.
LoGo has a sparser  and more significant inference network with  smaller fractions of false positives which stay below $5n$, which is anyway a large number of misclassification.
Nonetheless, we observe that, despite such large fractions of $\FP$, the discovered true positives are statistically significant.

\begin{table}[!h]
\caption{
{\bf Unconditioned transfer entropy network: comparison between fraction of true positive (TP/n) and fraction of false positive (FP/n).}
Statistical validation of  transfer entropy is at $p_v=1\%$ p-value. 
The table reports only the case for the parameter $\gamma=0.1$. 
}
\begin{center}
\begin{tabular}{@{\extracolsep{4pt}}lllllllll@{}}
\hline
q & 10 & 20 & 30 & 50 & 200 & 300 & 1000 & 20000  \\ \hline 
ridge TP/n  & 0.02  & 0.39$^{**}$  & 0.45$^{**}$  & 0.51$^{**}$  & 0.65$^{**}$  & 0.69$^{**}$  & 0.78$^{**}$  & 0.92$^{**}$ \\ 
ridge FP/n  & 0.07  & 1.06  & 0.95  & 0.85  & 0.93  & 0.99  & 1.20  & 1.73 \\ \hline 
Glasso TP/n  & 0.00  & 0.24$^{**}$  & 0.35$^{**}$  & 0.43$^{**}$  & 0.57$^{**}$  & 0.60$^{**}$  & 0.67$^{**}$  & 0.77$^{**}$ \\ 
Glasso FP/n  & 0.00  & 0.10  & 0.20  & 0.29  & 0.51  & 0.56  & 0.73  & 1.66 \\ \hline 
LoGo TP/n  & 0.11  & 0.34$^{**}$  & 0.41$^{**}$  & 0.47$^{**}$  & 0.63$^{**}$  & 0.66$^{**}$  & 0.76$^{**}$  & 0.89$^{**}$ \\ 
LoGo FP/n  & 0.02  & 0.16  & 0.25  & 0.34  & 0.59  & 0.66  & 0.87  & 1.49 \\ \hline 
&& & $^{**}$  $P < 10^{-8}$ &&&\\
\end{tabular}\end{center}
\label{tab:fraction_Unconditional}
\end{table}

\subsection{Unconditioned transfer entropy network}
We last tested whether conditioning to {the past} of all other variables gives better causality network retrievals than the unconditioned case. 
Here, transfer entropy, $T(\mathbf{ Z_i} \rightarrow \mathbf{ Z_j})$, is  computed by using  Eq.\ref{TE20} with  $\mathbf W=\emptyset$, the empty set.
For the ridge case this unconditional transfer entropy depends only from the time-series, $\mathbf{ Z}_{i,t}$, $\{\mathbf{ Z}_{i,t-1},...,\mathbf{ Z}_{i,t-\tau}\}$ and $\{\mathbf{ Z}_{j,t-1},...,\mathbf{ Z}_{j,t-\tau}\}$ (with $\tau = 5$ in this case).
Glasso and LoGo cases are instead hybrid because a conditional dependency has been already introduced in the sparse structure of the inverse covariance $\mathbf J$ (the inference network).
Results are reported in Tab.\ref{tab:fraction_Unconditional} where we observe that these networks retrieve a larger quantity of true positives than the ones constructed from conditional entropy. 
However, the fraction of false positive is also  larger than the ones in Tab.\ref{tab:fraction} although it is smaller than what observed in the inference network in Tab.\ref{tab:fraction_Inference}.
{Overall,   these results indicate that conditioning is effective in discarding false positives. }

% \begin{center}
%	\begin{figure}[t]
%%	\includegraphics[width=0.45\textwidth]{ROC_all.pdf}
%%	\includegraphics[width=0.45\textwidth]{ROC_all_detail.pdf}
%		\caption{
%		{\bf ROC values, for each model and each parameter combination.} X-axis  false positive rates  ($\FP/m$), y-axis true positive rates ($\TP/n$).
%		Left and right figures are the same with X-axis expanded on the low values only for the right figure to better visualise the differences between the various models.
%		Large symbols refer to $\gamma = 0.1$ and  validation at p-value $p_v=0.01$.
%		Color intensity is proportional to time series length.
%		Inference network results are  all outside the range of the plot. 
%		Reported values are averages over 100 processes.	}
%		\label{f.ROC}
%	\end{figure}
%\end{center}

\subsection{Summary of all results in a single ROC plot}
In summary, we have investigated the networks associated with conditional transfer entropy, unconditional  transfer entropy and inference for three models under a range of different parameters.
In the previous sub-sections we have provided some comparisons between the performances of the three models in different ranges of  parameters.
Let us here provide a summary of all results within a single ROC plot \cite{swets2014signal}.
Figure \ref{f.ROC}  reports the ROC values, for each model and each parameter combination, x-axis  are false positive rates  ($\FP/m$) and y-axis true positive rates ($\TP/n$).
Each point is an average over 100 processes.
Points above the diagonal line are associated to relatively well performing models with the upper left corner representing the point where models correctly discover all true causality links without any false positive.
The plot reports with large symbols the cases for $\gamma = 0.1$ and validation at p-value $p_v=0.01$, which can be compared with the data reported in the tables.
\CR{We note that, by construction, LoGo models are sparse (with a number of edges $\sim 3p$ \cite{LoGo16}). This restrains the ROC results to the left-hand side of the plot.
For this reason an expanded view of the figure is also proposed with the x-axis scaled.
Note that this ROC curve is provided as a visual  tool for intuitive comparison between models.}

Overall from Tables \ref{tab:fraction}, \ref{tab:fraction_Inference}, \ref{tab:fraction_Unconditional} and Fig. \ref{f.ROC} we conclude that all models obtain better results for longer time series and that conditional transfer entropy over-performs the unconditional counterparts (see, Tables  \ref{tab:fraction} and  \ref{tab:fraction_Unconditional} and  the two separated ROC figures for conditional and unconditional transfer entropies reported in Fig.\ref{f.ROCCoUc} in appendix \ref{A.Co.UC}).
In the range of short time series, when $q\le p$, which is of interest for this paper,  LoGo is the best performing model with better performances achieved for small $\gamma \lesssim 10^{-4}$ and validation with  small p-values $p_v \lesssim 10^{-4}$.
LoGo is consistently the best performing model also for longer time-series up to lengths of  $q \sim1000$.
Instead, above $q=2000$ ridge begins to provide better results.
For long time series, at  $q=20,000$, the best performing model is ridge with  parameters $\gamma =10^{-5}$,  p-value $p_v = 5\;10^{-6}$.
LoGo is also performing well when time series are long with best performance obtained at $q=20,000$ for parameters $\gamma =10^{-10}$,  p-value $p_v = 5\;10^{-6}$.
We note that LoGo instead performs  poorly in the region of parameters with $\gamma \le 0.1$ and $p_v \le 0.01$ for short time-series $q\le p/2$.

\section{Conclusions and perspectives}\label{s.conclusions}

In this paper we have undertaken the challenging task to explore models and parameter regions where the analytics of time series can retrieve significant fractions of true causality links from linear multivariate autoregressive process with known causality structure.
Results demonstrate that sparse models with conditional transfer entropy are the ones who achieve best results with significant causality link retrievals  already for very short time series even with $q \le p/5 = 20$.
\CR{This region is very critical and general considerations would suggest that no solutions can be discovered.
Indeed, this result is in apparent contradiction with a general analytical results in \cite{varga2016replica,papp2016fluctuation} who find that no significant solutions should be retrieved for $q \le N/2 = 150$.
However, we notice that the problem we are addressing here is different from the one in \cite{varga2016replica,papp2016fluctuation}.
In this paper we have been considering an underlying sparse true causality structure and such a sparsity changes considerably the condition of the problem yielding to significant solutions even well below the theoretical limit from \cite{varga2016replica,papp2016fluctuation}  which is instead associated to non-sparse models.
}

Unexpectedly, we observed that the structure of the inference networks in the two sparse models, Glasso and LoGo, have excessive numbers of false positives yielding to rather poor performances. 
However, in these models false positive can be efficiently filtered out by imposing statistical significance of the transfer entropies. 

Results are affected by the choice of the parameters and the fact that the models depend on various parameters ($q$, $p$, $\gamma$,  $p_v$, $P$) make the navigation in this space quite complex.
We observed that the choice of  p-values, $p_v$, for valid transfer entropies affects results.
Within our setting we obtained best results with the smaller p-values especially in the regions of short time-series.
We note that the regularizer parameter $\gamma$ also plays an important role and best performances are obtained by combination of the two parameters $\gamma$ and $p_v$. 
Not surprizingly, longer time-series yield to better results.
We observe that conditioning  to all other variables or unconditioning is affecting the transfer entropy estimation with better performing causality network retrieval obtained for conditioned transfer entropies.
However, qualitatively, results are comparable. Other intermediate cases, such as conditioning to past of all other variables only, have been explored again with qualitatively comparable results.  
It must be said that in the  present system results are expected to be robust to different conditionings because the underlying network of the investigated processes is sparse. For denser inference structures, conditioning could affect more the results.

Consistently with the findings in \cite{LoGo16} we find that LoGo outperforms the other methods.
This is encouraging because the  present settings of LoGo is using a simple class of information filtering networks, namely the TMFG \cite{TMFG}, obtained by retaining largest correlations.
There are a number of alternative information filtering networks which should be explored. % \cite{}.
In particular, given the importance of statistical validation emerged from the present work, it would be interesting to explore statistical validation within the process of construction of  the information filtering networks themselves.

In this paper we investigate a simple case with a linear autoregressive multivariate normal process analysed by means of linear models. 
Both LoGo and Glasso can be extended to the non-linear case with LoGo being particularly  suitable for non-parametric approaches as well \cite{LoGo16}.

There are Alternative methods to extract causality networks from short time series, in particular  Multispatial CCM \cite{sugihara2012detecting,clark2015spatial}  appears to perform well for short time series.
A comparison between different approaches and the application of these methods to real data will be extremely interesting.
However this should be the object of future works.

\subsection*{Acknowledgement} 
 T.A. acknowledges support of the UK Economic and Social Research Council (ESRC) in funding the Systemic Risk Centre (ES/K002309/1). 
 TDM wishes to thank the COST Action TD1210 for partially supporting this work and Complexity Science Hub Vienna. 

\subsection*{Conflict of interest disclosure} 
The authors declare that there is no conflict of interest regarding the publication of this paper.

\appendix

\section{Conditional transfer entropy}\label{cTE}
\label{s.definitions1}
Let us here briefly review two of the most commonly used information theoretic quantities, that we  use in this paper, namely, mutual information (quantifying dependency) and transfer entropy (quantifying causality) for the multivariate case \cite{shannon2001mathematical,schreiber2000measuring,anderson1984multivariate}.

\subsection{Mutual information}
Let us first start from  the simplest case of two random variables, $X\in \mathbb R^{1}$ and $Y\in \mathbb R^{1}$, where dependence can be quantified by the amount of shared information between the two variables, which is called mutual information: $I(X;Y)=H(X)+H(Y)-H(X,Y)$ where $H(X)$ is the entropy of variable $X$, $H(Y)$ is the entropy of variable $Y$ and $H(X,Y)$ is the joint entropy of variables $X$ and $Y$ \cite{anderson1984multivariate}.
Extending to the multivariate case, the shared information between a set of $n$ random variables $\mathbf X = (X_1,...,X_n)^T\in \mathbb R^{n}$ and another set of $m$ random variables $\mathbf Y = (Y_1,...,Y_m)^T \in \mathbb R^{m}$ is 
\begin{align} \label{I1}
I(\mathbf X;\mathbf Y) & {} 
 = H(\mathbf X) + H(\mathbf Y) - H(\mathbf X,\mathbf Y) 
\end{align}
with $H(\mathbf X)$, $H(\mathbf Y)$,  the  entropies respectively for the set of variables $\mathbf X$ and $\mathbf Y$ and $H(\mathbf X,\mathbf Y)$ their joint entropy.
It must be stressed that this quantity is the mutual information between two sets of multivariate variables and it is not the multivariate mutual information between all variables $\{ \mathbf X,\mathbf Y \}$ which instead measures the intersection of information between all variables. 
Mutual information  in Eq.\ref{I1} can also be written as
\begin{align}\label{I2}
I(\mathbf X;\mathbf Y) & {} 
= H(\mathbf Y) - H(\mathbf Y|\mathbf X)=H(\mathbf X) - H(\mathbf X|\mathbf Y)
\end{align}
which makes use of  the conditional entropy of $\mathbf Y$ given $\mathbf X$:  $H(\mathbf Y|\mathbf X)=H(\mathbf Y,\mathbf X)-H(\mathbf X)=\mathbb E(H(\mathbf Y)|\mathbf X)$.

Conditioning to a third set of variables $\mathbf W$ can also be applied to mutual information itself and its expression is a direct extension of Eq.\ref{I1} and it is called conditional mutual information:
\begin{align}\label{Ic}
I(\mathbf X;\mathbf Y|\mathbf W) & {} 
 = H(\mathbf X|\mathbf W) + H(\mathbf Y|\mathbf W) - H(\mathbf X,\mathbf Y|\mathbf W) \;\;.
\end{align}
Eq.\ref{I1} and Eq.\ref{Ic} coincide in the case of an empty set $\mathbf W=\emptyset$.
Mutual information and conditional mutual information are symmetric measures with $I(\mathbf X;\mathbf Y|\mathbf W) =I(\mathbf Y;\mathbf X|\mathbf W) $ always.
Let us note that symmetry is unavoidable for information measures that quantify the simultaneous effect of a set of variables onto another. 
Indeed, in a simultaneous interaction cause and effect cannot be distinguished from the exchange of information  and  direction cannot be established.
To quantify causality  one must investigate the transmission of information not only between two sets of variables but also trough  time. %

\subsection{Conditional transfer entropy}\label{cTE.s}
Causality between two random variables, $X\in \mathbb R^{1}$ and $Y\in \mathbb R^{1}$, can be quantified by means of the so-called transfer entropy which quantifies the amount of uncertainty on $Y$ explained by {the past} of $ X $ given {the past} of  $Y$. Let us consider a series of observations  and denote with $X_{t}$ the random variable $X$ at time $t$ and with $X_{t-\tau}$ the random variable at a previous time, $\tau$ lags before $t$. Using this notation, we can define transfer entropy from variable $X$ to variable $Y$ in terms of the following conditional mutual information: $T(X \rightarrow Y) = I(Y_{t};X_{t-\tau}|Y_{t-\tau})$ \cite{schreiber2000measuring,anderson1984multivariate}.

For the multivariate case, given two sets of random variables $\mathbf{ X} \in \mathbb R^{n}$ and $\mathbf{ Y} \in \mathbb R^{m}$, the transfer entropy  is the conditional mutual information between the set of variables $\mathbf{Y}_t$ at time $t$ and {the past} of the other set of variables, $\mathbf{ X}_{t-\tau}$ conditioned to {the past} of  the first variable $\mathbf{Y}_{t-\tau}$.
 This is:
$T(\mathbf X \rightarrow \mathbf Y) = I(\mathbf{Y}_t;\mathbf{ X}_{t-\tau}|\mathbf{ Y}_{t-\tau}) $ \cite{anderson1984multivariate}.
In general, the influence from {the past} can come from more than one lag and we can therefore extend the  definition including different sets of lags for the two variables: $\tau_1,...,\tau_k$, $\lambda_1,...,\lambda_h$:
\begin{align}\label{GeneralTE}
T(\mathbf X \rightarrow \mathbf Y) & {}= I(\mathbf{Y}_t; \{\mathbf{ X}_{t-\tau_1}...\mathbf{ X}_{t-\tau_k}\}|\{\mathbf{ Y}_{t-\lambda_1}...\mathbf{ Y}_{t-\lambda_h} \}) \\%\;\;,
& {} = H(\mathbf{Y}_t | \{\mathbf{ Y}_{t-\lambda_1}...\mathbf{ Y}_{t-\lambda_h} \}) - H(\mathbf{Y}_t | \{\mathbf{ X}_{t-\tau_1}...\mathbf{ X}_{t-\tau_k},\mathbf{ Y}_{t-\lambda_1}...\mathbf{ Y}_{t-\lambda_h} \}) \nonumber
\end{align}
a further generalization, which we use in this paper, includes conditioning to any other set of variables  $\{\mathbf{ W}_{t-\theta_1}...\mathbf{ W}_{t-\theta_g}\}$ lagged at $\theta_1,...,\theta_g$:
\begin{align}\label{General_cTE1}
T(\mathbf X \rightarrow \mathbf Y|\mathbf{ W}) & {}= I(\mathbf{Y}_t; \{\mathbf{ X}_{t-\tau_1}...\mathbf{ X}_{t-\tau_k}\}|\{\mathbf{ Y}_{t-\lambda_1}...\mathbf{ Y}_{t-\lambda_h},\mathbf{ W}_{t-\theta_1}...\mathbf{ W}_{t-\theta_g}\}) \;\;.
\end{align}

In this paper we simplify notation using $\mathbf{ X}_{t}^{lag}=\{\mathbf{ X}_{t-\tau_1}...\mathbf{ X}_{t-\tau_k}\}$,  $\mathbf{ Y}_{t}^{lag} = \{\mathbf{ Y}_{t-\lambda_1}...\mathbf{ Y}_{t-\lambda_h}\}$ and $\mathbf{ W}_{t} = \{\mathbf{ W}_{t-\theta_1}...\mathbf{ W}_{t-\theta_g}\}$.

In the literature, there are several examples that use adaptations of Eq.\ref{General_cTE} to compute causality and dependency measures \cite{pearl2009causality}.
A notable example is the directed information, introduced by Massey in \cite{massey1990causality},  where $\tau$ spans all lags in a range between $0$ to $s-1$,  $\lambda$ spans the lags from $1$ to $s-1$. 
The  directed information is then defined as the sum over transfer entropies from $s=1$ to present:
\begin{align}
I(\{\mathbf X\}_{1}^t \rightarrow \{\mathbf Y\}_{1}^t|\mathbf{ W}) & {}= \sum_{s=1}^t I(\mathbf{Y}_s; \{\mathbf{ X}\}_{1}^s|\{\mathbf{ Y}\}_{1}^{s-1},\mathbf{ W}) \;\;.
\end{align}
where we adopted the notation $\{\mathbf X\}_{1}^t = \{\mathbf{ X}_{1}...\mathbf{ X}_{t}\}$ and $\{\mathbf Y\}_{1}^t = \{\mathbf{ Y}_{1}...\mathbf{ Y}_{t}\}$.
Interestingly, this definition includes the conditional synchronous mutual information contributions between $\mathbf{ X}_{s}$ and $\mathbf{ Y}_{s}$.
Following Kramer \cite{kramer1998directed,amblard2012relation} we observe that for stationary processes
\begin{align}
\lim_{t \to \infty} \frac{1}{t} I(\{\mathbf X\}_{1}^t \rightarrow \{\mathbf Y\}_{1}^t) = \lim_{t \to \infty}  I(\{\mathbf X\}_{1}^t; \mathbf Y_t | \{\mathbf Y\}_{1}^{t-1})=T(\{\mathbf X\}_{1}^{t-1} \rightarrow \mathbf Y_t) + I(\{\mathbf X\}_{1}^t ; \{\mathbf Y\}_{1}^t |\{\mathbf X\}_{1}^{t-1})\;\;,
\end{align}
with $T(\{\mathbf X\}_{1}^{t-1} \rightarrow \mathbf Y_t)= I(\mathbf{Y}_t; \{\mathbf{ X}_{1}...\mathbf{ X}_{t-1}\}|\{\mathbf{ Y}_{1}...\mathbf{ Y}_{t-1}\} )$.
This identity supports the intuition that the directed information accounts for the transfer entropy plus an instantaneous term. 

\section{Shannon-Gibbs entropy}\label{Shannon}
The general expression for the transfer entropy reported di in Sec.\ref{s.definitions1}, Eq.\ref{General_cTE} is independent on the kind of entropy definition.
In this paper we use Shannon entropy, which is defined as
\begin{align}\label{entropy}
H(\mathbf X)  & {}=  - \mathbb E [ \log p(\mathbf X) ] \\
H(\mathbf Y)  & {}=  - \mathbb E [ \log p(\mathbf Y) ] 
\end{align}
where $p(\mathbf X)$ and $p(\mathbf Y)$ are the probability distribution function for the set of  random variables $\mathbf X$ and  $\mathbf Y$. 
Similarly, the joint Shannon entropy for the variables $\mathbf X$ and $\mathbf Y$ is defined as
\begin{align}
H(\mathbf X,\mathbf Y) =  - \mathbb E [ \log p(\mathbf X,\mathbf Y) ] 
\end{align}
with $p(\mathbf X,\mathbf Y )$  the  joint probability distribution function of  $\mathbf X$ and  $\mathbf Y$. 
This is the most common definition of entropy. It is a particularly meaningful and suitable entropy for linear modelling, as we  focus in the  paper.

\subsection{Multivariate normal modelling} \label{ss.linearModelling}
For multivariate normal variables the Shannon-Gibbs entropy is: 
\begin{align}\label{HLoGo}
H(\mathbf X) =  \frac12 \log \left(\mbox{det} {\mathbf \Sigma(\mathbf X)}\right)  + \frac{n}2 \log\left( 2\pi e \right)
\end{align}
and its conditional counterpart is
\begin{align}\label{H2}
H(\mathbf X | \mathbf W) =  \frac12 \log\left( { \mbox{det} \mathbf \Sigma(\mathbf X | \mathbf W)} \right) + \frac{n}2 \log\left( 2\pi e \right)
\end{align}
with $\mathbf \Sigma$ the covariance matrix and $\mbox{det}( . )$ the matrix determinant. 
In the paper we use these expressions to compute mutual information and conditional transfer entropy.

\begin{center}
	\begin{figure}[t]
	\includegraphics[width=0.80\textwidth]{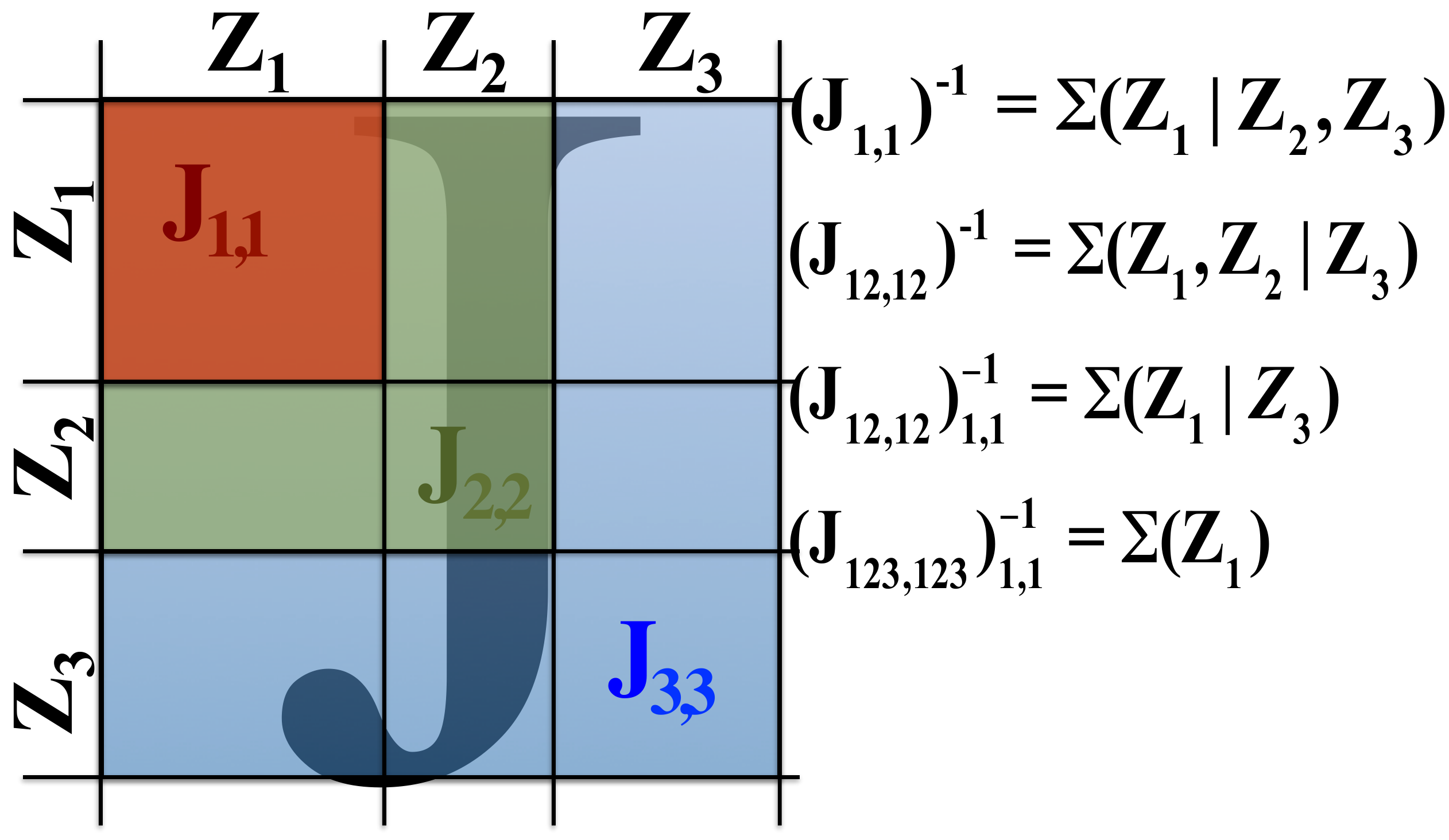}
		\caption{
		The inverse of parts the inverse covariance $\mathbf J$ gives the  covariance of the variables corresponding to that part conditioned to the other variables.
		}
		\label{f.inversions}
	\end{figure}
\end{center}

\section{ Computing conditional covariances for sub-sets of variables from the inverse covariance }\label{InverseCovJandSigma}

Let us consider three  sets of variables $\mathbf{ Z_1} \in \mathbb R^{p_1}$, $\mathbf{ Z_2} \in \mathbb R^{p_2}$ and $\mathbf{ Z_3} \in \mathbb R^{p_3}$ and  the associated  inverse covariance $\mathbf J \in \mathbb R^{{(p_1+p_2+p_3)}\times{(p_1+p_2+p_3)}}$ for  $\{\mathbf{ Z_1 },\mathbf{ Z_2 },\mathbf{ Z_3 }\} \in \mathbb R^{(p_1+p_2+p_3)}$.
The conditional covariance of $\mathbf{ Z_1}$ given $\mathbf{ Z_2}$ and $\mathbf{ Z_3}$ is  the inverse of the $p_1 \times p_1$ upper left part of $\mathbf J$ with indices in $V_1=(1,...,p_1)$ (see Fig.\ref{f.inversions}):
\begin{align}\label{Cov1_3}
 {\mathbf\Sigma}(\mathbf{Z_1}|\mathbf{Z_2},\mathbf{Z_3}) = \left(\mathbf{ J_{1,1}}\right)^{-1} \;\;.
\end{align}

Instead, the conditional covariance of $\mathbf{Z_1}$ given $\mathbf{Z_3}$ is obtained by inverting the larger upper left part  $\mathbf{ J_{12,12}}$ with both indices in $\{V_1,V_2\}$ with $V_2=(p_1+1,...,p_1+p_2)$, and then taking  the inverse of the part with indices in  $V_1$ which, using  the  Schur complement \cite{anderson1984multivariate}, is:
\begin{align}\label{Cov2_3}
 {\mathbf\Sigma}(\mathbf{Z_1}|\mathbf{Z_3}) =  (\mathbf {J_{1,1}} - \mathbf {J_{1,2}} (\mathbf {J_{2,2}})^{-1} \mathbf {J_{2,1}})^{-1} \;\;.
\end{align}

Figure \ref{f.inversions} schematically illustrates  these inversions and their relations with conditional covariances. 
Let us note that these conditional covariances can also be expressed directly in terms of sub-covariances by using again the Schur complement:
\begin{align}\label{Cov1_3COV}
 {\mathbf\Sigma}(\mathbf{Z_1}|\mathbf{Z_2},\mathbf{Z_3}) 
&= {\mathbf{ \Sigma_{1,1}}}- \mathbf{{ \Sigma_{1,23}}}( {\mathbf{ \Sigma_{23,23}}})^{-1} {\mathbf{ \Sigma_{23,1}}}\\
 {\mathbf\Sigma}(\mathbf{Z_1}|\mathbf{Z_3}) 
&= {\mathbf{ \Sigma_{1,1}}}- \mathbf{{ \Sigma_{1,3}}}( {\mathbf{ \Sigma_{3,3}}})^{-1} {\mathbf{ \Sigma_{3,1}}}\;\;.
\label{Cov1_3COV1}
\end{align}
However, when $p_3$ (cardinality of $V_3$) is much larger than $p_1$ and $p_2$ (cardinalities of $V_1$ and $V_2$)  then the equivalent expressions, Eqs.\ref{Cov1_3} and \ref{Cov2_3}, that use the  inverse covariance involve matrices with much smaller dimensions.
This can become computationally crucial when very large dimensionalities are involved. 
Furthermore, if the  inverse covariance $\mathbf J$ is estimated by using a sparse modeling tool such as Glasso or LoGo \cite{friedmanetal2008,LoGo16} (as we  do in this paper), then  computations in expressions Eqs.\ref{Cov1_3} and \ref{Cov2_3} have to handle only a few non-zero elements providing great computational advantages over  Eqs.\ref{Cov1_3COV} and \ref{Cov1_3COV1}.

In the paper we  make use of Eqs.\ref{Cov1_3}-\ref{Cov2_3} to compute mutual information and conditional transfer entropy for the system of all variables and their lagged versions. 

\subsection{Mutual information}
Let us consider the mutual information between any two subsets $\mathbf X \in \mathbb R^n$ and $\mathbf Y \in \mathbb R^m$ of variables conditioned to all other variables, which we shall call $\mathbf W\in \mathbb R^{p-n-m}$.
For these three sets of variables  $\{\mathbf X,\mathbf Y,\mathbf W\} \in \mathbb R^{p}$ the conditional mutual information, $I(\mathbf X,\mathbf Y,\mathbf W)=H(\mathbf X,\mathbf Y|\mathbf W)-H(\mathbf X|\mathbf Y,\mathbf W)$ (Eq.\ref{Ic}), can be expressed in terms of the conditional covariances by using Eq.\ref{H2}:
\begin{align}\label{JMI0c}
I(\mathbf X;\mathbf Y|\mathbf W) = \frac12 \log \mbox{det} {\mathbf\Sigma}(\mathbf X|\mathbf W)- \frac12 \log  \mbox{det} {\mathbf\Sigma}(\mathbf X | \mathbf Y,\mathbf W) \;\;.
\end{align}
Given the   inverse covariance $\mathbf J \in \mathbb R^{p\times p}$, by using  Eqs.\ref{Cov1_3} and \ref{Cov2_3} and substituting 
\begin{align}
\mathbf{Z_1} &= \mathbf X\\ \nonumber
\mathbf{Z_2} &= \mathbf Y\\ \nonumber
\mathbf{Z_3} &= \mathbf W
 \end{align}
 we can express the  conditional mutual information, Eq.\ref{JMI0c}, directly in terms of the parts of $\mathbf J$: 
\begin{align}\label{JMI1c}
I(\mathbf X;\mathbf Y|\mathbf W) =  - \frac12 \log \mbox{det}\left( \mathbf {J_{1,1}} - \mathbf {J_{1,2}} (\mathbf {J_{2,2}})^{-1} \mathbf {J_{2,1}})\right)+ \frac12 \log \left (\mbox{det}\mathbf{ J_{1,1} }\right ) \;\;.
\end{align}
Note that, although this is not directly evident, Eq.\ref{JMI1c} is symmetric by exchanging $\mathbf 1$ and $\mathbf 2$ (i.e. $\mathbf X$ and $\mathbf Y$).

\subsection{Conditional transfer entropy }\label{s.CTE}
Conditional transfer entropy (Eq.\ref{General_cTE} ) is a conditional mutual information between lagged sets of variables and  therefore it can be computed directly from Eq.\ref{JMI1c}.
In this case we shall  name
\begin{align}\label{TE20b}
\mathbf{Z_1} &= \mathbf{Y}_t\\ \nonumber
\mathbf{Z_2} &=\{\mathbf{ X}_{t-\tau_1}...\mathbf{ X}_{t-\tau_k}\}\\ \nonumber
\mathbf{Z_3} &=  \{\mathbf{ Y}_{t-\lambda_1}...\mathbf{ Y}_{t-\lambda_h},\mathbf{ W}_{t-\theta_1}...\mathbf{ W}_{t-\theta_g}\}\\\nonumber
T(\mathbf X \rightarrow \mathbf Y|\mathbf{ W})  &=  - \frac12 \log \mbox{det} \left(\mathbf {J_{1,1}} - \mathbf {J_{1,2}} (\mathbf {J_{2,2}})^{-1} \mathbf {J_{2,1}})\right)+ \frac12 \log \mbox{det} \left (\mathbf{ J_{1,1} }\right )
 \end{align}
obtaining an expression which is formally identical to Eq.\ref{JMI1c} but with indices $\mathbf{ 1}$ and $ \mathbf 2$ referring to the above sets of variables instead.

Note that the index $\mathbf 3$ does not appear in this expression. 
Information from variables  $\mathbf 3$ ($\mathbf W$) has been used to compute $\mathbf J$ but then only the sub-parts   $\mathbf{ 1}$ and $ \mathbf 2$ are required to compute the conditional transfer entropy. 
The fact that these expressions for conditional mutual information and conditional transfer entropy involve only local parts ($\mathbf{ 1}$ and $ \mathbf 2$) of the inverse covariance can become extremely useful when high dimensional datasets are involved.

\begin{center}
	\begin{figure}[t]
	\includegraphics[width=0.45\textwidth]{ROC_cond.pdf}
	\includegraphics[width=0.45\textwidth]{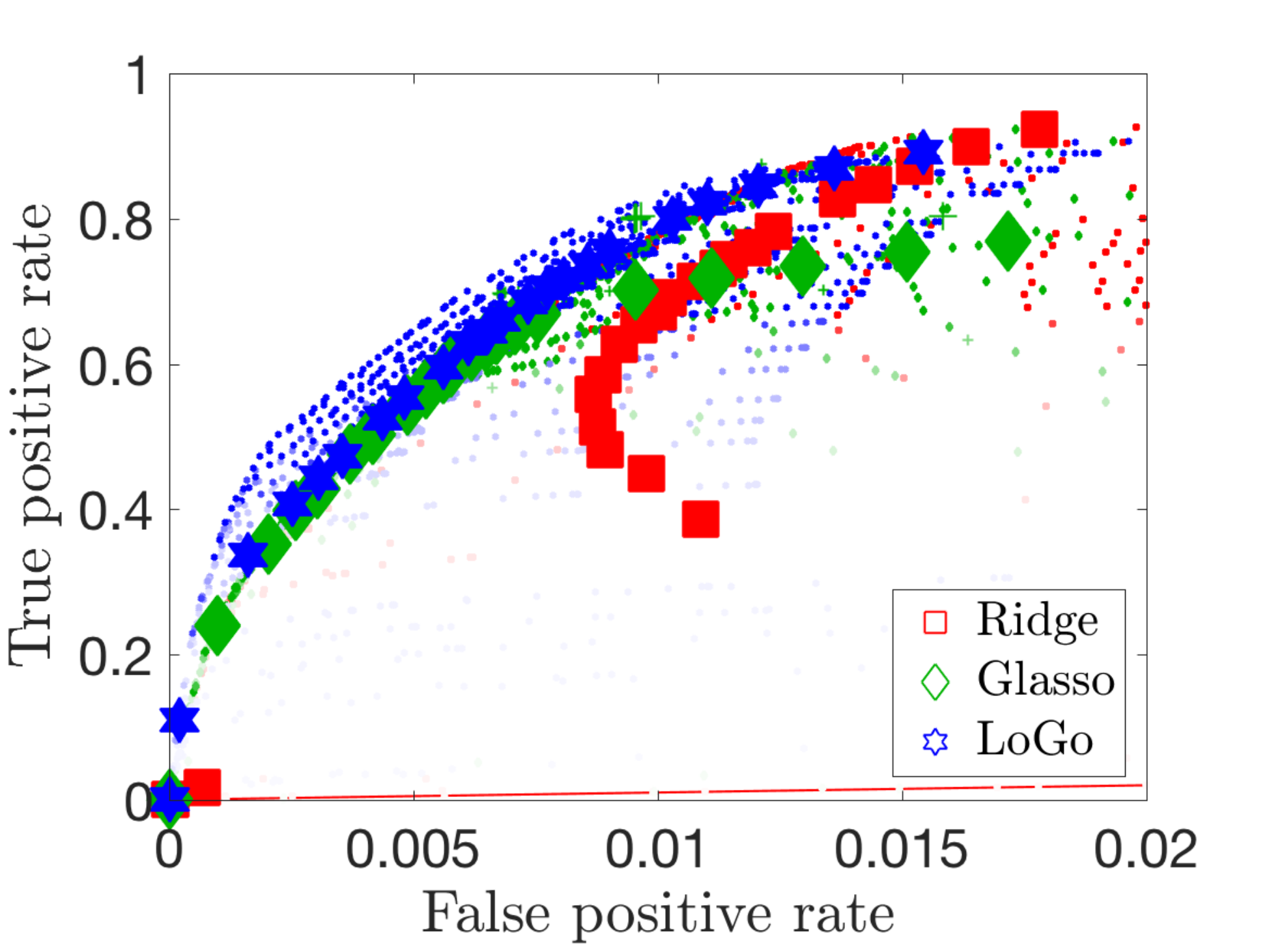}
		\caption{
		{\bf ROC values, for conditional (left) and unconditional (right) transfer entropies.} X-axis  false positive rates  ($\FP/m$), y-axis true positive rates ($\TP/n$).
		Large symbols refer to $\gamma = 0.1$ and  validation at p-value $p_v=0.01$.
		Color intensity is proportional to time series length.
		Inference network results are  all outside the range of the plot. 
				Reported values are averages over 100 processes.	
		\label{f.ROCCoUc}}
	\end{figure}
\end{center}

\section{Comparison between conditional and unconditional transfer entropies}\label{A.Co.UC}
The two ROC plots for conditional and unconditional transfer entropies are displayed in Fig.\ref{f.ROCCoUc}. 
Form the comparison it is evident that, for the  process studied in this paper, conditional transfer entropy provides best results.
This is in line with what observed in Tab.s~\ref{tab:fraction},\ref{tab:fraction_Unconditional},\ref{tab:fraction_B} and \ref{tab:fraction_U_B}.

\section{Causality network results for transfer entropy validation with 1\% Bonferroni adjusted p-values} \label{BonferroniValidated}
In tables \ref{tab:fraction_B} and  \ref{tab:fraction_U_B}, are reported true positive rates (${\TP/n}$) and fraction of false  positives ($\FP/m$) statistically validated, causality links with validation at 1\% Bonberroni adjusted p-value (i.e. $p_v \lesssim 10^{-6}$). These tables must be compared with  Tab.s \ref{tab:fraction} and  \ref{tab:fraction_Unconditional},  in the main text where causality links are validated at $p_v=1\%$ non-adjusted p-value.

\begin{table}[!h]
\caption{
{\bf Causality network validation with conditional transfer entropy validation at 1\% Bonberroni adjusted p-value.} 
Fraction of true positive (${TP/n}$) and  fraction of false  positive  ($FP/n$), statistically validated, causality links for the three models and different time-series lengths.
The table reports only the case for the parameter $\gamma=0.1$. 
}
\begin{center}
\begin{tabular}{@{\extracolsep{4pt}}lllllllll@{}}
\hline
q & 10 & 20 & 30 & 50 & 200 & 300 & 1000 & 20000  \\ \hline 
ridge TP/n  & 0.00  & 0.00  & 0.00  & 0.00  & 0.00  & 0.30$^{**}$  & 0.67$^{**}$  & 0.89$^{**}$ \\ 
ridge FP/n  & 0.00  & 0.00  & 0.00  & 0.00  & 0.00  & 0.01  & 0.18  & 0.75 \\ \hline 
Glasso TP/n  & 0.00  & 0.00  & 0.00  & 0.00  & 0.35$^{**}$  & 0.43$^{**}$  & 0.57$^{**}$  & 0.71$^{**}$ \\ 
Glasso FP/n  & 0.00  & 0.00  & 0.00  & 0.00  & 0.01  & 0.03  & 0.13  & 0.45 \\ \hline 
LoGo TP/n  & 0.00  & 0.00  & 0.02  & 0.17$^{**}$  & 0.50$^{**}$  & 0.56$^{**}$  & 0.69$^{**}$  & 0.87$^{**}$ \\ 
LoGo FP/n  & 0.00  & 0.00  & 0.00  & 0.00  & 0.01  & 0.03  & 0.09  & 0.28 \\ \hline 
&& &$^{**}$  $P < 10^{-8}$ &&&\\
\end{tabular}\end{center}
\label{tab:fraction_B}
\end{table}

\begin{table}[!h]
\caption{
{\bf Causality network validation with unconditional transfer entropy validation at 1\% Bonberroni adjusted p-value.} 
Fraction of true positive  (${TP/n}$) and fraction of false  positive ($FP/n$), statistically validated, causality links for the three models and different time-series lengths.
The table reports only the case for the parameter $\gamma=0.1$. 
}
\begin{center}
\begin{tabular}{@{\extracolsep{4pt}}lllllllll@{}}
\hline
q & 10 & 20 & 30 & 50 & 200 & 300 & 1000 & 20000  \\ \hline 
ridge TP/n  & 0.00  & 0.00  & 0.22$^{**}$  & 0.36$^{**}$  & 0.55$^{**}$  & 0.59$^{**}$  & 0.70$^{**}$  & 0.88$^{**}$ \\ 
ridge FP/n  & 0.00  & 0.00  & 0.09  & 0.21  & 0.47  & 0.55  & 0.77  & 1.32 \\ \hline 
Glasso TP/n  & 0.00  & 0.00  & 0.00  & 0.27$^{**}$  & 0.48$^{**}$  & 0.53$^{**}$  & 0.62$^{**}$  & 0.75$^{**}$ \\ 
Glasso FP/n  & 0.00  & 0.00  & 0.00  & 0.11  & 0.37  & 0.43  & 0.61  & 1.41 \\ \hline 
LoGo TP/n  & 0.00  & 0.00  & 0.22$^{**}$  & 0.35$^{**}$  & 0.53$^{**}$  & 0.58$^{**}$  & 0.69$^{**}$  & 0.86$^{**}$ \\ 
LoGo FP/n  & 0.00  & 0.00  & 0.05  & 0.16  & 0.42  & 0.49  & 0.71  & 1.27 \\ \hline 
&& &$^{**}$  $P < 10^{-8}$ &&&\\
\end{tabular}\end{center}
\label{tab:fraction_U_B}
\end{table}

\bibliographystyle{unsrt}
%\bibliography{logo_CausalNet}

\end{document}